\documentclass{article}
\usepackage{spconf,amsmath,amssymb,graphicx}
\graphicspath{{figures/}}

\newcommand{\figref}[1]{Fig.~\ref{#1}}
\newcommand{\tbref}[1]{Table~\ref{#1}}

\title{JOINT GROUP TESTING OF TIME-VARYING FAULTY SENSORS AND SYSTEM STATE ESTIMATION IN LARGE SENSOR NETWORKS}
\name{Mengqi Ren and Ruixin Niu}
\address{Department of Electrical and Computer Engineering\\
Virginia Commonwealth University\\
Richmond, Virginia 23284, U.S.A.}
\begin{document}
%
\maketitle
\begin{abstract}
The problem of faulty sensor detection is investigated in large sensor networks where the sensor faults are sparse and time-varying, such as those caused by attacks launched by an adversary. Group testing and the Kalman filter are designed jointly to perform real time system state estimation and time-varying faulty sensor detection with a small number of tests. Numerical results show that the faulty sensors are efficiently detected and removed, and the system state estimation performance is significantly improved via the proposed method. Compared with an approach that tests sensors one by one, the proposed approach reduces the number of tests significantly while maintaining a similar fault detection performance. 

\end{abstract}
\section{Introduction}
\label{sec:intro}

Sensor faults happen when sensors return corrupted data \cite{Harvey, Zhou, Xue}. The corruption could be caused by attacks from an adversary, sensor malfunctioning, or disturbance from the environment. In a multi-sensor system, detecting faulty sensors is crucial to ensure the system's normal operation, since system state estimates based on faulty sensor measurements are misleading. 

In many cases, the faulty sensors are sparse in sensor networks. For example, if sensors are attacked by an adversary, typically only a small number of sensors are attacked and corrupted due to the adversary's limited resources and his/her intention to reduce the chance of being detected by the system defender. Furthermore, the adversary may adopt a time-varying attack strategy to further reduce the probability of being detected. If the attacks are sparse, it is not necessary to test all the sensor nodes which could be laborious and inefficient in a large system. So our aim is to reliably detect/identify the sparse faulty sensors, and at the same time to significantly reduce the costs associated with testing the sensors. One promising approach that could achieve this goal is group testing \cite{Atia}. It is a well known search method and can be viewed as a Boolean version of compressive sensing \cite{Donoho, Baraniuk, Candes}, where the sparse vector only consists of binary entries and Boolean matrix multiplication is used to generate compressed testing results which contain all the information of the sparse vector.

There is little work on fault/failure detection using group testing in dynamic systems with time-varying fault states. One related publication is  \cite{Lo2013}, in which a fault detection method based on combinatorial group testing and the Kalman filter was proposed. In this method, each testing group is divided into two subgroups and two Kalman filters are run separately on them. The detection decision for each testing group is made by comparing predicted state estimates of the two Kalman filters.  
Note that in \cite{Lo2013}, only the problem of {\it time-invariant} faulty sensor detection was investigated. The problem of sparse fault/failure detection in distributed sensor networks was studied  in \cite{Tosic}. To reduce communication cost, the group testing procedure is successively separated into two phases, in which all the sensors only need to communicate with their neighbors. However, this method requires the fault state to be {\it time-invariant} while preforming group testing over phases. In addition, all the above mentioned approaches perform only sensor fault/failure detection but not system state estimation.

Typically, faulty sensor detection and system state estimation are realized separately and the latter is implemented after all the faulty sensors are detected and removed from the system. To detect the faulty sensor(s) via group testing, a time consuming optimization problem needs to be solved. Therefore, it is difficult to implement this procedure in real-time systems. In this paper, a new approach for joint group testing of time-varying faulty sensors and system state estimation is proposed, in which system state is estimated before decoding the fault state of sensors so that the system state can be estimated in real time. A modified group testing is developed to realize the time-varying faulty sensor detection. 


\section{Problem Formulation} 
\label{sec:probform}

In this paper, linear dynamic systems are considered. The system state could be modeled by the following discrete-time linear system state equation \cite{YBS:book}
\begin{equation}
\label{eq:state}
\mathbf{x}_{k+1} = \mathbf{Fx}_k + \mathbf{\Gamma}\mathbf{v}_k
\end{equation}
where $\mathbf{x}_k$ is the $n_x \times 1$ state vector at time $k$, $\mathbf{F}$ is $n_x \times n_x$ state transition matrix, $\mathbf{v}_k$ is the process noise at time $k$, and $\mathbf{\Gamma}$ is the gain matrix for $\mathbf{v}_k$. Furthermore, $\{\mathbf{v}_k\}$ is a sequence of white Gaussian process noise with $E(\mathbf{v}_k)=\mathbf{0}$ and $E(\mathbf{v}_k \mathbf{v}_k^T)=\mathbf{Q}_k$ for all $k=0,1,2,\dots$.

Let us consider a large sensor network which is composed of $N$ sensors. Denote this sensor network as a set ${\cal N}=\{1,2,\dots,N\}$. Assume that only a few sensors in the sensor network are corrupted by adversary and the fault state of sensors is time-varying. Denote the set of faulty sensors at time $k$ as ${\cal D}_k$ which is a subset of ${\cal N}$. The components of ${\cal D}_k$ are time-varying as different sensors are attacked over time. Denote the size of ${\cal D}_k$ by $D_k$, which is also time-varying and $D_k \ll N$. To detect faulty sensors, the state of each sensor is represented by two hypotheses $H_0$ and $H_1$. Let us assume that under hypothesis $H_0$, sensor $i$ is normal, and  its measurement equation is
\begin{equation}
\mathbf{z}^i_k = \mathbf{H}^i\mathbf{x}_k + \mathbf{w}^i_k
\end{equation}
where $\mathbf{z}^i_k$ is the $n_z \times 1$ measurement vector of sensor $i$ at time $k$, $\mathbf{H}^i$ is the $n_z \times n_x$ measurement matrix of sensor $i$, and $\mathbf{w}^i_k$ is the measurement noise of sensor $i$ at time $k$. Also, $\{\mathbf{w}^i_k\}$ is a sequence of white Gaussian measurement noise with $E(\mathbf{w}^i_k)=\mathbf{0}$ and $E(\mathbf{w}^i_k {\mathbf{w}^i_k}^T)=\mathbf{R}^i_w$ for $k=1,2,\dots$ and $i=1,2,\dots,N$.

Under hypothesis $H_1$, sensor $i$ is faulty and its  measurement equation is
\begin{equation}
\mathbf{z}^i_k = \mathbf{H}^i\mathbf{x}_k + \mathbf{w}^i_k + \mathbf{b}^i_k
\end{equation}
where $\mathbf{b}^i_k$ is the bias vector which is injected by the adversary to sensor $i$ at time $k$.

The Kalman filter is used to process the sensor measurements. To maintain the performance of the Kalman filter, the measurements of time-varying faulty sensors should be removed adaptively. This motivates joint group testing of time-varying faulty sensors and system state estimation.

\section{Joint Time-Varying Fault Detection and System State Estimation}
\label{sec:JTFDT}

Since a large sensor network is considered and the faulty sensors are assumed to be sparse in the sensor network, the group testing is adopted to detect sensor faults. Group testing implements tests on several testing groups which are generated by binary probabilistic sampling matrix, and the indicator vector of defective sensors is decoded from the testing results. Typically, group testing is applied at each point in time. In this paper, a new group testing structure is designed over a period of time. By doing this, the fault detection method is able to detect faulty sensors when their quantity and indices are time-varying. Meanwhile, the number of tests and computation costs can be reduced significantly. 

The fault state of all the sensors during a time period $K$ is indicated by a $KN$-dimensional binary vector $\mathbf{f} \in \text{GF}^{KN}(2)$, where $\text{GF}(2)$ is a Galois field of order two \cite{golan2012linear}. The $\mathbf{f}(i)=1$ indicates sensor $1+[(i-1) \text{ mod } N]$ at time $\lceil i/N \rceil$ is faulty whereas $\mathbf{f}(i)=0$ indicates a normal sensor. Denote the sparsity level of $\mathbf{f}$ by $d$, and clearly $d=\sum_{k=1}^{K}D_k$. Assume $T$ testing groups are generated in total. The tests performed on the sensor network are represented with $T \times KN$ probabilistic sampling matrix $\mathbf{\Phi}$. If $\mathbf{\Phi}(i,j)=1$, then the sensor $1+[(j-1) \text{ mod } N]$ at time $\lceil j/N \rceil$ is selected in the $i$th testing group. The entries of $\mathbf{\Phi}$ follow i.i.d. Bernoulli($p$). In noise-free model, group testing outcome vector $\mathbf{g}$ is obtained as follows
\begin{equation}
\label{eq:GT_noise-free}
\mathbf{g} = \mathbf{\Phi} \odot \mathbf{f}
\end{equation}
where $\mathbf{g} \in \text{GF}^T(2)$, and $\odot$ denotes the Boolean matrix multiplication operator which is composed of the logical AND and OR operators.
In the presence of noise, group testing results are inverted which can be illustrated by the following simple model \cite{Malioutov}
\begin{equation}
\label{eq:GT_noisy}
\mathbf{g} = (\mathbf{\Phi} \odot \mathbf{f}) \oplus \mathbf{e}
\end{equation}
where $\oplus$ denotes XOR operator, $\mathbf{e} \in \text{GF}^T(2)$ is the Boolean vector of errors which represents the effect of noise. The ones in $\mathbf{e}$ indicate corruption and they will invert the corresponding results of $\mathbf{\Phi} \odot \mathbf{f}$ which leads to false alarms or misses. Note that the model \eqref{eq:GT_noisy} is one simple way to illustrate the presence of noise/mistake and $\mathbf{e}$ is difficult to model  in some cases.

\begin{figure}[htb]
	\centering
	\includegraphics[width=8.5cm]{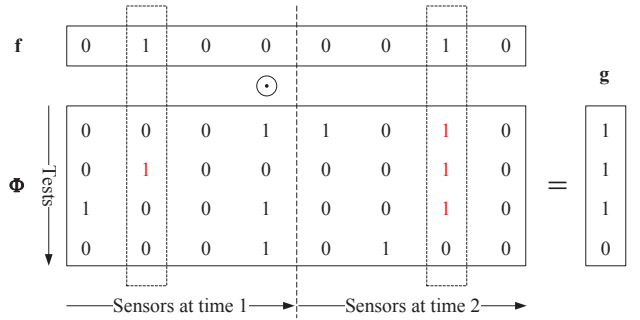}
	\caption{An example of time-varying group testing problem}
	\label{fig:GT_time-varying}
\end{figure}

A toy example of the noise-free group testing procedure is shown in \figref{fig:GT_time-varying}. In this example, $N=4$, $K=2$, and $D_1=D_2=1$. The 2nd sensor at time 1 and the 3rd sensor at time 2 are faulty. The 2nd sensor at time 1 is selected in test 2 and the 3rd sensor at time 2 is selected in tests 1, 2, and 3. As long as one faulty sensor is selected in a specified test, the outcome of this test will be 1. If no faulty sensor is selected in a test, then the testing outcome is 0. Therefore, the outcomes of test 1, 2, and 3 are ones and the outcome of test 4 is zero.

To decode the fault state vector $\mathbf{f}$ efficiently, the probabilistic sampling matrix $\boldsymbol{\Phi}$ should satisfy $d$-disjunct property as it ensures identifiability of the $d$-sparse fault state vector. A matrix $\boldsymbol{\Phi}$ is called $d$-disjunct if for any $d+1$ columns, there always exists a row with entry 1 in a column and zeros in all the other $d$ columns \cite{Atia}. In the example shown in \figref{fig:GT_time-varying}, $\boldsymbol{\Phi}$ is a 2-disjunct probabilistic sampling matrix. The fault state vector $\mathbf{f}$ is decoded via the linear programming (LP) relaxation as in \cite{Malioutov}, in which the inputs are $\mathbf{g}$ and $\mathbf{\Phi}$, and the output is $\mathbf{f}$.

Note that the outcome of group testing is a binary vector but the measurements are continuous. We need to find a way to decide whether a testing group contains faulty sensors or not. The innovation of Kalman filter is a good choice as it is a zero-mean, white, and Gaussian sequence. To achieve real-time system state estimation and detect time-varying faulty sensors, joint group testing of time-varying faulty sensors and system state estimation are proposed and described as the following steps.

{\bf Step I}: build testing groups. Generate $T \times KN$ probabilistic sampling matrix $\boldsymbol{\Phi}$ via Bernoulli($p$). Let us divide $\boldsymbol{\Phi}$ into $K$ blocks by column, where each block is $T \times N$ sub-matrix $\boldsymbol{\Phi}_k$. Denote the $t$-th testing group, i.e. $t$-th row, in $\boldsymbol{\Phi}_k$ by ${\cal G}_{t,k}$. The size of ${\cal G}_{t,k}$ is denoted by $G_{t,k}$,where $1 \leqslant t \leqslant T$ and $1 \leqslant k \leqslant K$.

{\bf Step II}: generate outcome vector $\mathbf{g}$. Run Kalman filter from $k=1$ to $K$. For each time $k$, run Kalman filter by using each testing group ${\cal G}_{t,k}$, then obtain innovation $\boldsymbol{\nu}_{t,k}$ and measurement prediction covariance $\mathbf{S}_{t,k}$. If all the sensors in ${\cal G}_{t,k}$ are normal, then $\boldsymbol{\nu}_{t,k}$ is a zero-mean Gaussian random variable and it can be tested via $\chi^2$ test: $\boldsymbol{\nu}_{t,k}^T\mathbf{S}_{t,k}^{-1}\boldsymbol{\nu}_{t,k} \sim \chi^2(n_z G_{t,k})$. Moreover, the innovation is a white sequence if no faulty sensor in ${\cal G}_{t,k}$, and we will have the following distribution
\begin{equation}
\label{eq:Chi2}
\sum_{s=1}^{k} \boldsymbol{\nu}_{t,s}^T\mathbf{S}_{t,s}^{-1}\boldsymbol{\nu}_{t,s} \sim \chi^2(n_z\sum_{s=1}^{k}G_{t,s})
\end{equation}
Note that if ${\cal G}_{t,k}=\emptyset$, do not run Kalman filter in test $t$ at time $k$ and skip the corresponding item in \eqref{eq:Chi2}. The outcome vector $\mathbf{g}$ is generated via \eqref{eq:Chi2} as follows: If \eqref{eq:Chi2} is satisfied at time $k$, the next innovation $\boldsymbol{\nu}_{t,k+1}$ is calculated and tested. If all the testing groups ${\cal G}_{t,1},{\cal G}_{t,2},\dots,{\cal G}_{t,K}$ in test $t$ satisfy \eqref{eq:Chi2}, the outcome of test $t$ is negative and $g_t = 0$. Otherwise, this procedure is stopped for test $t$ as long as \eqref{eq:Chi2} is not satisfied,  the outcome of test $t$ is positive, and $g_t = 1$. In this way, not all testing groups chosen by $\boldsymbol{\Phi}$ are fully tested as this procedure may stop when $k < K$, which saves computational costs. 

{\bf Step III}: tracking object via Kalman filter. At each time $k$, test all the sensor groups ${\cal G}_{1,k},{\cal G}_{2,k},\dots,{\cal G}_{T,k}$ via \eqref{eq:Chi2}. Form a normal sensor group by taking the union of all the sensor groups which pass the test. Run Kalman filter on this normal sensor group, then we obtain updated state estimate and updated state covariance, which are used as the inputs of Kalman filter in both Step II and Step III at the next time step, and as the system state estimate output of the algorithm. 

{\bf Step IV}: identify faulty sensors via group testing. The fault state vector $\mathbf{f}$ is decoded by solving the LP relaxation as in \cite{Malioutov}.

Note that  system state estimation is implemented before decoding $\mathbf{f}$ which is time consuming. This design guarantees real-time system state estimation with faulty sensors in large sensor networks.

As we mentioned before, the aim of group testing is reducing the number of tests. Here we derive the  upper bound on the average  number of $\chi^2$ tests required by the proposed method. According to Step II, the $\chi^2$ test in test $t$ at time $k$ is skipped if ${\cal G}_{t,k}=\emptyset$. So, one upper bound is the number of nonempty sets ${\cal G}_{t,k}$ among the $T \times K$ tests. Since the entries of $\mathbf{\Phi}$ follow i.i.d. Bernoulli($p$), the probability of ${\cal G}_{t,k}$ being nonempty is $1-(1-p)^N$. Therefore, the  upper bound on the average number of $\chi^2$ tests in designed group testing is $TK[1-(1-p)^N]$. 

If the sensors are tested via $\chi^2$ test one by one at each time, we can design a similar testing procedure. The only differences are $T=N$ and $G_{t,k}=1$ for all $t \in \{1,2,\dots,T\}$ and $k \in \{1,2,\dots,K\}$. The number of $\chi^2$ tests in the one-by-one testing approach is $KN$. 

\section{Simulation Results}
\label{sec:Simulation}

For simplicity we give a multi-sensor target tracking example to illustrate the effectiveness of the proposed approach. Let us assume that an object is moving in a 1-dimensional space with its state at time $k$ denoted by $\mathbf{x}_k=[\varphi_k\;\; \dot{\varphi}_k]^T$, where $\varphi_k$ and $\dot{\varphi}_k$ are the object's position and velocity at time $k$, respectively. The state transition matrix is
\[\mathbf{F}
= \left[\begin{array}{cc}
1& T_s\\
0& 1
\end{array}
\right]
\]
where $T_s=0.1$ seconds is the time interval between two measurements. The process noise gain matrix $\mathbf{\Gamma}$ in \eqref{eq:state} is $[{T_s^2}/{2}\;\; T_s]^T$. The variance of state process noise is $Q=0.01$. The mean and covariance matrix of the object's initial state are $\mathbf{\hat{x}}_{0|0}=[0\;\; 1.5]^T$ and $\mathbf{P}_{0|0}=diag([1000, 1])$, respectively. Assume that there are $N=150$ sensors, all of which measure the object's position over time. Therefore, the measurement matrix is $\mathbf{H}^i=[1\;\; 0]$ for all $i$. The covariance matrix of sensor measurement noise is $\mathbf{R}^i_w=1$ for all $i$. 

Assume that the adversary chooses  sensors to attack randomly via Bernoulli($q$) where $q=0.01$. The bias injected by the adversary follows i.i.d. Gaussian distribution $\mathbf{b}^i_k \sim  \; {\cal N}(\mathbf{0}, \mathbf{R}_b)$ for all $i$. Choose $K=5$ to design $\boldsymbol{\Phi}$ and the entries of $\boldsymbol{\Phi}$ follow i.i.d. Bernoulli($p$) where $p=\frac{1}{qKN}=0.01$ \cite{Atia}. The number of testing groups is $T=50$ which is ${\cal O}(qKN log(KN))$. Two-sided $\chi^2$ test with 0.001 significance level is applied in Step II in Section \ref{sec:JTFDT}. The regularization parameter in LP relaxation \cite{Malioutov} is set as 1. All the results are based on $100$ Monte Carlo simulations. 

\begin{figure}[htb]
	\centering
	\includegraphics[width=8.5cm]{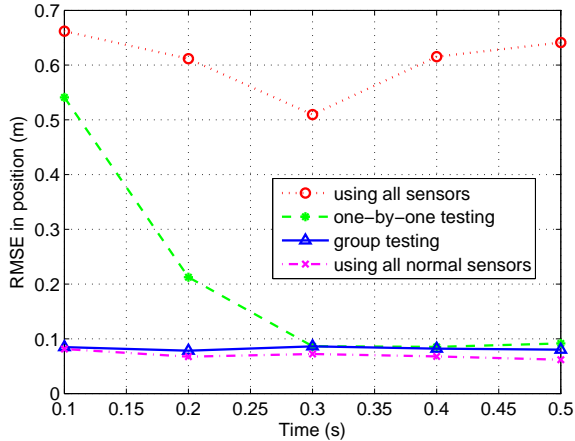}
	\caption{\small RMSE of Position over time}
	\label{fig:RMSE_Pos}
\end{figure}

\begin{figure}[htb]
	\centering
	\includegraphics[width=8.5cm]{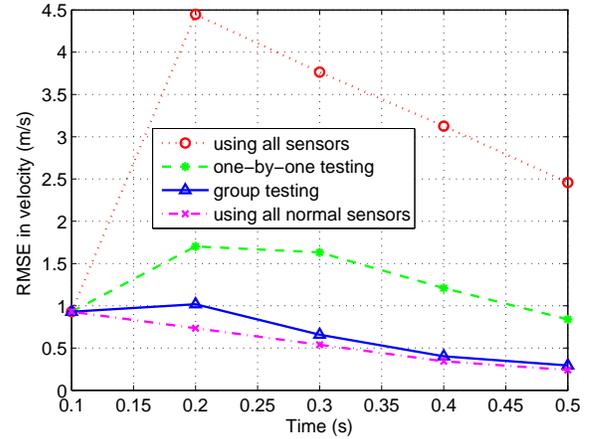}
	\caption{\small RMSE of Velocity over time}
	\label{fig:RMSE_Vel}
\end{figure}

To evaluate the tracking performance of the proposed method, it is compared with two methods: one is using the measurements from all the sensors, the other is testing all the sensors one by one and only uses the ones passes a $\chi^2$ test to track the target. The performance of the three methods is compared in terms of root mean squared error (RMSE) of position and velocity. Assume $\mathbf{R}_b=10000$ which means the injected bias noise by the adversary is strong. The simulation results are shown in Figs. \ref{fig:RMSE_Pos} and \ref{fig:RMSE_Vel}. It is clear that the RMSEs of position and velocity of the proposed method are the smallest among the three methods and they are close to the RMSEs achieved by a clairvoyant  Kalman filter using all the normal sensors. That is to say, the proposed method chooses normal sensors efficiently when tracking the object. The proposed method has better performance than testing all the sensors one by one since the degrees of freedom of the $\chi^2$ distributions in the proposed method are larger. Furthermore, both of the RMSE of position and velocity are small by using the proposed method, which means this method is robust to attacks with strong injection noise.

To study the fault detection performance of the proposed method under different levels of attacks,  $\mathbf{R}_b$ is changed from $100$ to $50000$ and probability of errors are evaluated under different $\mathbf{R}_b$. The simulation results are shown in \tbref{tb:Pe_Rb}, in which test$_1$ stands for one-by-one test and test$_2$ stands for group testing. The probabilities of false alarm of these two methods are very close. The probabilities of miss of these two methods are also close to each other when $\mathbf{R}_b$ is between $10^2$ and $10^4$. For both methods, probability of false alarm is smaller than probability of miss as the significance level of the $\chi^2$ test is low.
\begin{table}[htb]
\small
 	\caption{Probability of errors under different $\mathbf{R}_b$}
 	\begin{center}
 		\label{tb:Pe_Rb}
 		\begin{tabular}{|l|r|r|r|r|r|}
 			\hline
 			$\mathbf{R}_b$ & 100 & 1000 & 5000 & 10000 & 50000 \\
 			\hline
 			test$_1$ $P_{fa}$ & 0.0104 & 0.0007 & 0.0008 & 0.0039 & 0.0006 \\
 			\hline
 			test$_2$ $P_{fa}$ & 0.0056 & 0.0114 & 0.0115 & 0.0140 & 0.0155 \\
 			\hline
 			test$_1$ $P_{m}$ & 0.4629 & 0.2552 & 0.2058 & 0.1554 & 0.0800 \\
 			\hline
 			test$_2$ $P_{m}$ & 0.5027 & 0.3738 & 0.3380 & 0.3824 & 0.3535 \\
 			\hline
 		\end{tabular}
 	\end{center}
\end{table}

The average number of $\chi^2$ tests is shown in \tbref{tb:N_Chi2}, in which the theoretical upper bound on the average number of tests in group testing is shown in the second row. Clearly, the results are in accordance with the theoretical value and the average number of $\chi^2$ tests in designed group testing is about $\frac{1}{4}$ of the one-by-one test. Considering results of the second simulation, the proposed method is able to achieve similar fault detection performance to the one-by-one test by using a much smaller number of tests. 
\begin{table}[htb]
\small
 	\caption{Average number of $\chi^2$ tests under different $\mathbf{R}_b$}
 	\begin{center}
 		\label{tb:N_Chi2}
 		\begin{tabular}{|l|r|r|r|r|r|}
 			\hline
 			$\mathbf{R}_b$ & 100 & 1000 & 5000 & 10000 & 50000 \\
 			\hline
 			test$_1$ & 750 & 750 & 750 & 750 & 750 \\
 			\hline
 			upper bound & 250 & 250 & 250 & 250 & 250 \\
 			\hline
 			test$_2$ & 197 & 184 & 179 & 178 & 172 \\
 			\hline
 		\end{tabular}
 	\end{center}
\end{table}




\section{Conclusion}
\label{sec:Conclusion}
In this paper, a new approach for joint time-varying faulty sensor detection and system state estimation was proposed by combining group testing and Kalman filter. A new group testing structure was developed to detect the time-varying fault state of sensors. To realize real-time tracking, system state estimation is performed without the full knowledge of fault state of sensors. It was shown from simulations that the proposed method significantly improves the state estimation performance in the presence of faulty sensors and it has higher estimation performance than an alternative one-by-one test. Compared to the one-by-one test, the proposed method achieves similar fault detection performance by using a much smaller number of tests.

%


\vfill\pagebreak

\bibliographystyle{IEEEbib}
\bibliography{references}

\end{document}